\documentclass[aps, prd, twocolumn, showpacs, showkeys, superscriptaddress, preprintnumbers, floatfix]{revtex4}
\usepackage{multirow}
\usepackage{slashed}
\usepackage{graphicx}
\usepackage{amsmath}
\usepackage{bm}
\usepackage{yhmath}
\usepackage{hyperref}
\usepackage{subfigure}
\usepackage[T1]{fontenc}
\usepackage{color}
\usepackage{cases}
\usepackage{subfigure}
\usepackage{dcolumn, booktabs, bm}
\usepackage{amsfonts, amssymb, stmaryrd, latexsym, amsmath}
\usepackage{textcomp}
\usepackage{amsthm}
\usepackage{mathrsfs}

\usepackage{array}

\newcounter{RomanNumber}

\newcommand{\lyxmathsym}[1]{\ifmmode\begingroup\def\b@ld{bold}
	\text{\ifx\math@version\b@ld\bfseries\fi#1}\endgroup\else#1\fi}

\renewcommand{\arraystretch}{1.9}
\usepackage{braket}
\allowdisplaybreaks

\begin{document}
	\title{Masses of hidden-charm pentaquark states with $J^P = \frac{3}{2}^-$}
	
	\author{Tao Li}
	\email{tao@stumail.nwu.edu.cn}
	\affiliation{School of Physics, Northwest University, Xi'an 710127, China}
	\author{Hao-Song Li}
	\email{haosongli@nwu.edu.cn}
	\affiliation{School of Physics, Northwest University, Xi'an 710127, China}
	\affiliation{Shaanxi Key Laboratory for Theoretical Physics Frontiers, Xi'an 710127, China}
	\affiliation{Peng Huanwu Center for Fundamental Theory, Xi'an 710127, China}
	\begin{abstract}
		Within the heavy pentaquark chiral perturbation theory, we calculate the chiral corrections for $J^P=\frac{3}{2}^-$ octet hidden-charm pentaquark masses up to next-to-leading order. Taking the LHCb-reported $P_{\psi}^N(4440)$ and $P_{\psi s}^{\Lambda}(4459)$ (with $J^P=\frac{3}{2}^-$) as inputs, we predict the other two octet hidden-charm pentaquark states \(P_{\psi s}^{\Sigma}(4483)\) and \(P_{\psi ss}^{N}(4490)\) with $J^P=\frac{3}{2}^-$. The results provide theoretical guidance for the further search of \(P_{\psi s}^{\Sigma}\) and \(P_{\psi ss}^{N}\) in experiments.
			\end{abstract}
		\maketitle
		\thispagestyle{empty}
Multi-quark exotic hadron states, which transcend the traditional quark model description, are significant and attractive to investigate since they are one of the effective ways to deepen our understanding of the non-perturbative behavior of strong interactions~\cite{Brambilla:2010cs,Esposito:2016noz,Ali:2017jda}. In 2015, the LHCb Collaboration reported the initial detection of two hidden-charm pentaquark state candidates, $P_{\psi}^N(4380)$ and $P_{\psi}^N(4450)$, in the $J/\psi p$ invariant mass spectrum of the $\Lambda_{b}^{0} \to J/\psi p K^{-}$ decay process~\cite{LHCb:2015yax}. Since then, pentaquark states have become one of the hotspots in the field of theory and experiment research. In 2019, with the updated data, the LHCb Collaboration reported a new state $P_{\psi}^N(4312)$ and revealed that $P_{\psi}^N(4450)$ can split into two narrow peaks, $P_{\psi}^N(4440)$ and $P_{\psi}^N(4450)$~\cite{LHCb:2019kea}. Their masses and widths are
	\begin{eqnarray*}
		m_{P_{\psi}^N(4312)}&=&4311.9 \pm 0.7^{ +6.8}_{-0.6}~\mathrm{MeV}, \\\Gamma_{P_{\psi}^N(4312)}&=&9.8 \pm 2.7 ^{ +3.7}_{-4.5}~\mathrm{MeV},\\
		m_{P_{\psi}^N(4440)}&=&4440.3 \pm 1.3 ^{+ 4.1}_{-4.7}~\mathrm{MeV}, \\ \Gamma_{P_{\psi}^N(4440)}&=&20.6 \pm 4.9^{+8.7}_{-10.1}~\mathrm{MeV},\\
		m_{P_{\psi}^N(4457)}&=&4457.3 \pm 0.6 ^{+ 4.1}_{-1.7}~\mathrm{MeV}, \\ \Gamma_{P_{\psi}^N(4457)}&=&6.4 \pm 2.0^{+5.7}_{-1.9}~\mathrm{MeV}.
	\end{eqnarray*}
	However, the evidence for $P_{\psi}^N(4380)$ is neither confirmed nor refuted in the new discussion. $P_{\psi}^N(4312)$, $P_{\psi}^N(4440)$, and $P_{\psi}^N(4457)$ are very close to the $\Sigma_c\bar{D}$ and $\Sigma_c\bar{D}^*$ thresholds, so the molecular picture becomes a natural interpretation~\cite{Wang:2019ato,Chen:2019asm,Xiao:2019aya,Xiao:2019mvs,Guo:2019fdo,Guo:2019kdc,Burns:2019iih,Wang:2019spc}. In the molecular picture, the spin-parity quantum number $J^P=\frac{1}{2}^-$ for $P_{\psi}^N(4312)$ is universally accepted. However, for $P_{\psi}^N(4440)$ and $P_{\psi}^N(4457)$, there are two scenarios: $P_{\psi}^N(4440)$ and $P_{\psi}^N(4457)$ are the $\Sigma_c\bar{D}^*$ molecules with $J^P=\frac{1}{2}^-$ and $J^P=\frac{3}{2}^-$~\cite{He:2019ify,He:2019rva,Lin:2019qiv}, respectively (scenario A), or $J^P=\frac{3}{2}^-$ and $J^P=\frac{1}{2}^-$ for scenario B~\cite{Yalikun:2021bfm,PavonValderrama:2019nbk,Du:2021fmf,Peng:2024yzn}. Subsequently, the LHCb Collaboration announced a new pentaquark structure $P_{\psi}^N(4337)$ in $B_{s}\to J/\psi p\bar{p}$ decays~\cite{LHCb:2021chn} and predicted it with $J^{P} = 1/2^{+}$~\cite{Shen:2017ayv}. In 2020, the LHCb Collaboration observed the hidden-charm pentaquark state $P_{\psi s}^{\Lambda}(4459)$ containing s quark in the $J/\psi \Lambda$ invariant mass spectrum of the $\Xi_b^- \to J/\psi \Lambda K^-$ process~\cite{LHCb:2020jpq} with the mass and width
	\begin{eqnarray*}
		m_{P_{\psi s}^{\Lambda}(4459)}&=&4458.8 \pm 2.9^{+4.7}_{-1.1}~\mathrm{MeV}, \\ \Gamma_{P_{\psi s}^{\Lambda}(4459)}&=&17.3 \pm 6.5^{+8.0}_{-5.7}~\mathrm{MeV}.
	\end{eqnarray*}
	The $P_{\psi s}^{\Lambda}(4459)$ is close to the $\bar{D}^{*0}\Xi_c^0$ threshold, so it is naturally interpreted as a $\bar{D}^{*0}\Xi_c^0$ molecular state with \(J^P = 1/2^-\) or \(3/2^-\)~\cite{Zou:2021sha,Karliner:2021xnq,Peng:2020hql,Chen:2020uif,Chen:2020kco,Lu:2021irg}.
	Recently, $P_{\psi s}^{\Lambda}(4338)$ has been reported in the $J/\psi \Lambda$ invariant mass spectrum of the $B^-\to J/\psi \Lambda \bar{p}$ process and its amplitude analysis preferred the $J^P=\frac{1}{2}^-$~\cite{LHCb:2022ogu}. 
	
For the hidden-charm pentaquark states observed, enormous efforts are made to elucidate the quantum numbers of these states and their internal structure , with the most popular being the molecular state interpretation. In addition, compact pentaquark states~\cite{Ali:2019npk,Wang:2019got,Cheng:2019obk,Zhu:2019iwm,Pimikov:2019dyr,Ruangyoo:2021aoi}, virtual states~\cite{Fernandez-Ramirez:2019koa}, triangular singularities~\cite{Nakamura:2021qvy}, and cusp effects~\cite{Burns:2022uiv} are also taken up. It's critical to study the properties of these states systematically due to their uncertain nature. The particle quality provides new perspectives for further understanding of these states. Mass corrections are useful in more accurately determining whether the observed signals in experiments are new hidden-charm pentaquark states or interference from other known particles, thus helping to confirm  new discoveries. The hidden-charm pentaquark state, an exotic hadron, consists of four quarks and one antiquark. With mass corrections, we can explore the specific manifestations of the interquark color force in different quark combinations, refine theoretical descriptions of the hadronic internal structure, and further understand the quark model.

Chiral perturbation theory (ChPT), quantum chromodynamics (QCD) effective field theory at energies below the chiral symmetry breaking scale $\Lambda_{\chi} \sim 1 \text{ GeV}$~\cite{Weinberg:1978kz,Gasser:1983yg,Scherer:2012zzd,Huang:2019not}, is widely applied in modern investigations of low-energy hadron physics. It follows the QCD symmetry by using perturbative calculations at the hadron level instead of non-perturbative calculations at the quark level, and the obtained results show well-analytic. For the pentaquark mass calculation, we develop the heavy pentaquark chiral perturbation theory(HPChPT)~\cite{Li:2024jlq,Li:2025hpd}, which considers pentaquarks as extremely heavy to act as static sources, similar to the heavy baryon chiral perturbation theory(HBChPT)~\cite{Sun:2014aya,Jenkins:1990jv,Jenkins:1992pi}.

In this paper, we work in the framework of HPChPT, taking the observed octet hidden-charm pentaquark states $P_{\psi}^N(4440)$ and $P_{\psi s}^{\Lambda}(4459)$ with quantum numbers $J^P=\frac{3}{2}^-$ as inputs (see Refs \cite{Burns:2019iih,Yamaguchi:2019seo,Peng:2024yzn,Liu:2019zvb,Chen:2020kco,Chen:2021tip,Peng:2020hql,Wang:2022gfb,Ozdem:2024rch,Mutuk:2024ltc,Ozdem:2022kei}) to predict the octet hidden-charm pentaquark state masses $P_{\psi s}^{\Sigma}$ and $P_{\psi ss}^{N}$ with $J^P=\frac{3}{2}^-$.

To calculate the chiral correction of the mass with spin-$\frac{3}{2}$ pentaquark states, we need to construct the corresponding chiral effective Lagrangians. We first introduce the pseudoscalar meson field $\phi$, given by
\begin{equation}
	\phi=\left(\begin{array}{ccc}
		\pi^{0}+\frac{1}{\sqrt{3}}\eta & \sqrt{2}\pi^{+} & \sqrt{2}K^{+}\\
		\sqrt{2}\pi^{-} & -\pi^{0}+\frac{1}{\sqrt{3}}\eta & \sqrt{2}K^{0}\\
		\sqrt{2}K^{-} & \sqrt{2}\bar{K}^{0} & -\frac{2}{\sqrt{3}}\eta
	\end{array}\right).\nonumber
\end{equation}
The octet hidden-charm molecular pentaquark states $P_n$ reads
	\begin{equation}
	\setlength{\arraycolsep}{0.1pt}
	P_n=
	\left(		
	\begin{array}{ccc}
		\frac{1}{\sqrt{2}}{{}P_{\psi s}^{\Sigma}}^{0}+\frac{1}{\sqrt{6}}{{}P_{\psi s}^{\Lambda}}^{0}
		&{{}P_{\psi s}^{\Sigma}}^{+}
		&{{}P_{\psi}^{N}}^{+}
		\\
		{{}P_{\psi s}^{\Sigma}}^{-}
		&-\frac{1}{\sqrt{2}}{{}P_{\psi s}^{\Sigma}}^{0}+\frac{1}{\sqrt{6}}{{}P_{\psi s}^{\Lambda}}^{0}
		&{{}P_{\psi}^{N}}^{0}
		\\
		{{}P_{\psi ss}^{N}}^{-}
		&{{}P_{\psi ss}^{N}}^{+}
		&\frac{2}{\sqrt{3}}{{}P_{\psi s}^{\Lambda}}^{0}
		\nonumber\\
	\end{array}
	\right),
\end{equation} 
where n= 1,2 represent the $8_1$ and $8_2$ flavor pentaquark states, respectively. Under SU(3) symmetry, the $8_{1,2}$ flavor pentaquark state wave functions are shown in Ref.~\cite{Li:2024wxr}.

 The spin-$\frac{3}{2}$ free pentaquark state Lagrangian can be expressed as  
 \begin{eqnarray}
 	\mathcal{L}_0^{(1)} = - \text{Tr}[\bar{P}^\nu (iv \cdot D) P_\nu].
 \end{eqnarray}
 
 The covariant derivative acting on the pentaquark state is defined as
 \begin{eqnarray}
 	D_\mu &=& \partial_\mu + \frac{1}{2} (u^\dagger \partial_\mu u + u \partial_\mu u^\dagger),\\
 	u^2 &=& U = \exp(i\phi/F_\phi),
 \end{eqnarray}
 where $F_\phi$ is the Goldstone boson decay constant, $F_\pi\approx92.4$ MeV, $F_K\approx113$ MeV and $F_\eta\approx116$ MeV. 
  
  The second order pseudoscalar boson and octet pentaquark state interaction chiral Lagrangians read
  \begin{align}
  	\mathcal{L}_{\rm int1} &= 2g_3 \text{Tr} \left( \bar{P}_1^\nu S_\mu \{ u^\mu, P_{1\nu} \} \right) + 2f_3 \text{Tr} \left( \bar{P}_1^\nu S_\mu [ u^\mu, P_{1\nu} ] \right), \\
  	\mathcal{L}_{\rm int2} &= 2g_6 \text{Tr} \left( \bar{P}_2^\nu S_\mu \{ u^\mu, P_{2\nu} \} \right) + 2f_6 \text{Tr} \left( \bar{P}_2^\nu S_\mu [ u^\mu, P_{2\nu} ] \right),
  \end{align}
  where $g_{3,6}$ and $f_{3,6}$ are coupling constants and $S_\mu$ is a covariant spin operator. The so-called vielbein $u^\mu$ is defined as 
  \begin{eqnarray}
  	u^\mu = \frac{1}{2} i \left[ u^\dagger \partial_\mu u - u \partial_\mu u^\dagger \right].
  \end{eqnarray}

  We also need the leading order chiral effective Lagrangians contributing to the mass correction at the tree level
  \begin{eqnarray}
  	\mathcal{L}_1^{(2)} &=& h_1 \mathrm{Tr}\bigl(\bar{P}_1^\nu \{\chi_+, P_{1\nu}\}\bigr) + h_2 \mathrm{Tr}\bigl(\bar{P}_1^\nu [\chi_+, P_{1\nu}]\bigr)\nonumber \\
  	&&+h_3 \mathrm{Tr}(\bar{P}_1^\nu P_{1\nu})\mathrm{Tr}(\chi_+),\label{Eq:mass1}\\
  	\mathcal{L}_2^{(2)} &=& h_4 \mathrm{Tr}\bigl(\bar{P}_2^\nu \{\chi_+, P_{2\nu}\}\bigr) + h_5 \mathrm{Tr}\bigl(\bar{P}_2^\nu [\chi_+, P_{2\nu}]\bigr)\nonumber \\
  	&&+h_6\mathrm{Tr}(\bar{P}_2^\nu P_{2\nu})\mathrm{Tr}(\chi_+),\label{Eq:mass2}
  \end{eqnarray}
  where $h_i(i=1,\cdots,6)$ are the low-energy constant(LECs). The chiral block $\chi_+$ reads
\begin{eqnarray}
	\chi_{+}=u^\dag\chi u^\dag+ u\chi^\dag u,
\end{eqnarray}  
where $\chi=diag(M_{\pi}^2, M_{\pi}^2, 2M_K^2-M_{\pi}^2)$, the pseudoscalar meson masses $M_{\pi}=140$ MeV, $M_K=494$ MeV and $M_{\eta}=550$ MeV. 

The pentaquark states are considered as very heavy sources and equivalently their momentum decomposes as 
\begin{eqnarray}
	 p_\mu = m_0 v_\mu + k_\mu,
\end{eqnarray}
where $m_0$ is the pentaquark mass in the chiral limit, $v_\mu$ is the four-velocity constrained $v^2=1$, and $k_\mu$ is a small off-shell momentum satisfying $v\cdot k\ll m_0$.

\begin{figure}[htbp!]
\centering
\includegraphics[width=1\linewidth]{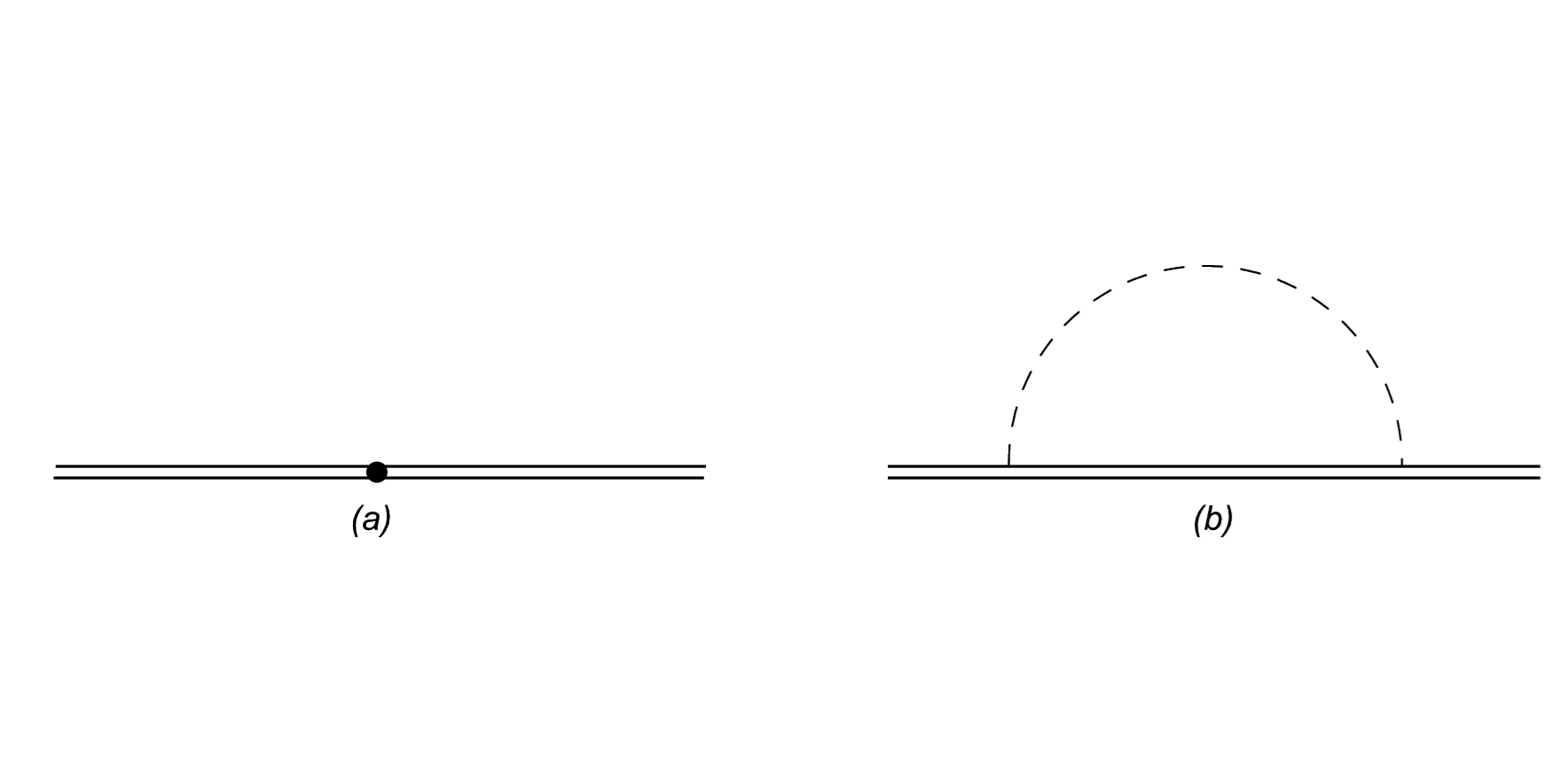}
\caption{Feynman diagrams contributing to the octet hidden-charm pentaquark state mass up to next-to-leading order. The double solid and dashed lines represent the octet hidden-charm pentaquark state and the Goldstone boson, respectively. Solid black dots represent second-order couplings.}
\label{fig:tree}
\end{figure}

The one-particle irreducible Feynman diagrams contributing to the mass corrections of octet hidden-charm pentaquarks with $J^P=\frac{3}{2}^-$ are shown in Fig.~\ref{fig:tree}. 

The corrections are given by the pentaquark self-energy  $\Sigma_P$, which is contained in the full propagator(in d dimensions)
\begin{eqnarray}
	G^{\rho\sigma}=\frac{-i P^{\rho\sigma}}{v\cdot p-m_0-\Sigma_P(\omega)},
\end{eqnarray}
where $\omega=v\cdot k$ and $P^{\rho\sigma} = g^{\rho\sigma}-v^\rho v^\sigma + 4\left(\frac{d-3}{d-1}\right)S^\rho S^\sigma$. The projection operator $P^{\rho\sigma}$ satisfies $v_\rho P^{\rho\sigma}=P^{\rho\sigma}v_\sigma=0$ ~\cite{Severt:2019sbz}. The physical pentaquark mass $m_P$ is the propagator pole at p = mv,
\begin{eqnarray}
	m_P = m_0 + \Sigma_{P}(0).
\end{eqnarray}

At $O(p^2)$, the tree level contribution corresponding to Fig.~\ref{fig:tree}(a) can be calculated from the chiral symmetry breaking Lagrangians in Eq.(~\ref{Eq:mass1}) and Eq.(~\ref{Eq:mass2}). At $O(p^3)$, the loop level in Fig.~\ref{fig:tree}(b) contributes to the self-energy
\begin{eqnarray}
	\Sigma_{P_{1,2}\phi}^{(b)} =-C_{P_{1,2}\phi}^{(b)} \frac{M_{\phi}^3}{16\pi F_{\phi}^2},
\end{eqnarray}
where the coefficients $C_{P_{1,2},\phi}^{(b)}$ are given in Table~\ref{tab_1}.
\begin{table}[htp]
	\centering
	\caption{Coefficients $C^{(b)}_{P_{1,2}\phi}$}
	\vspace{0.1em}
	\label{tab_1} 
	\renewcommand{\arraystretch}{2.5}
	\resizebox{82mm}{!}{
		\begin{tabular}{ c|c|c|c|c}
			\toprule[1.0pt]
			\toprule[1.0pt]
			States & ${P_{\psi}^{N}}$&${P_{\psi s}^{\Sigma}}$&${P_{\psi s}^{\Lambda0}}$&${P_{\psi ss}^{N}}$\\ \hline
			$C^{(b)}_{P_1\pi}$&$3(f_3+g_3)^2$&$\frac{4}{3}(6f_3^2+g_3^2)$&$4g_3^2$&$3(f_3-g_3)^2$ \\ \hline
			$C^{(b)}_{P_1K}$&$\frac{2}{3}(9f_3^2-6f_3g_3+5g_3^2)$&$4(f_3^2+g_3^2)$&$\frac{4}{3}(9f_3^2+g_3^2)$&$\frac{2}{3}(9f_3^2+6f_3g_3+5g_3^2)$\\\hline
			$C^{(b)}_{P_1\eta}$&$\frac{1}{3}(g_3-3f_3)^2$&$\frac{4}{3}g_3^2$&$\frac{4}{3}g_3^2$&$\frac{1}{3}(3f_3+g_3)^2$\\ \hline 
			$C^{(b)}_{P_2\pi}$&$3(f_6+g_6)^2$&$\frac{4}{3}(6f_6^2+g_6^2)$&$4g_6^2$&$3(f_6-g_6)^2$ \\ \hline
			$C^{(b)}_{P_2K}$&$\frac{2}{3}(9f_6^2-6f_6g_6+5g_6^2)$&$4(f_6^2+g_6^2)$&$\frac{4}{3}(9f_6^2+g_6^2)$&$\frac{2}{3}(9f_6^2+6f_6g_6+5g_6^2)$\\\hline
			$C^{(b)}_{P_2\eta}$&$\frac{1}{3}(g_6-3f_6)^2$&$\frac{4}{3}g_6^2$&$\frac{4}{3}g_6^2$&$\frac{1}{3}(3f_6+g_6)^2$\\
			\toprule[1.0pt]
			\toprule[1.0pt]
			
	\end{tabular}}
\end{table}	

The $8_1$ flavor hidden-charm pentaquark physical masses up to next-to-leading order(NLO) are expressed explicitly as
\begin{eqnarray}
	\label{equ:20}
	m_{P_{\psi}^{N}}&=&m_0+2h_3(M_\pi^2+2M_K^2)+4h_{2}(M_\pi^2-M_K^2)\nonumber\\
&&+4h_{1}M_K^2-\frac{3(f_3+g_3)^2 M_{\pi}^3}{16\pi F_\pi^2}
	-\frac{(g_3-3f_3)^2M_{\eta}^3}{48\pi F_{\eta}^2}\nonumber\\ 
	&&-\frac{(9f_3^2-6f_3g_3+5g_3^2)M_K^3}{24\pi F_K^2},		\\
	\nonumber\\
	\label{equ:21}
	m_{P_{\psi s}^{\Sigma}}&=&m_0+2h_3(M_\pi^2+2M_K^2)+4h_{1}M_\pi^2-\frac{g_3^2M_{\eta}^3}{12\pi F_{\eta}^2}\nonumber\\
	&&	-\frac{(6f_3^2+g_3^2)M_\pi ^3}{12\pi F_\pi^2}
	-\frac{(f_3^2+g_3^2) M_{K}^3}{4\pi F_K^2},
	\\
	\nonumber\\
	\label{equ:22}
	m_{P_{\psi s}^{\Lambda}}
	&=&m_0+2h_3(M_\pi^2+2M_K^2)+\frac{4}{3}h_{1}(4M_K^2-M_\pi^2)\nonumber\\
	&&-\frac{g_3^2M_\pi ^3}{4\pi F_\pi^2}
	-\frac{(9f_3^2+g_3^2) M_{K}^3}{12\pi F_K^2}
	-\frac{g_3^2M_{\eta}^3}{12\pi F_{\eta}^2},
	\\
	\nonumber\\
	\label{equ:23}
	m_{P_{\psi ss}^{N}}&=&m_0+2h_3(M_\pi^2+2M_K^2)+4h_{2}(M_K^2-M_\pi^2)\nonumber\\
	&&+4h_{1}M_K^2-\frac{3(f_3-g_3)^2 M_{\pi}^3}{16\pi F_\pi^2}
	-\frac{(3f_3+g_3)^2M_{\eta}^3}{48\pi F_{\eta}^2}\nonumber\\
	&&-\frac{(9f_3^2+6f_3g_3+5g_3^2)M_K^3}{24\pi F_K^2} .
\end{eqnarray}

To obtain the octet hidden-charm pentaquark state masses in HPChPT, we need to be determine six LECs: $m_0$, $h_1$, $h_2$, $h_3$, $f_3$, $g_3$, where $h_3$ can be absorbed into the bare mass $m_0$. In Ref.~\cite{Li:2024wxr}, we associated pentaquark states with nucleon axial charges using the quark model based on Lagrangians respecting symmetry. We obtained $f_3 = \frac{11}{90}g_A$, $g_3 = -\frac{1}{90}g_A$ and $f_6 = g_6 = \frac{1}{10}g_A$ via symmetry verification, here $g_A $ is the nucleon's axial charge. 

The three remaining unknown LECs, $m_0$, $h_1$ and $h_2$, need to be determined from the corresponding experimental information. Within the molecular $\bar{D}^*\Sigma_c$ description, assigning $J^P=\frac{3}{2}^-$ to $P_\psi ^N(4440)$ finds a positive effective range $r_0$ of natural size $O(1/\beta)$, where $1/\beta$ signifies the interaction range in Ref.~\cite{Peng:2024yzn}. The spin-parity of \(P_{\psi_s}^\Lambda(4459)\) state is preferably \(\frac{3}{2}^-\)~\cite{Mutuk:2024ltc}. We suppose $P_{\psi}^N(4440)$ and $P_{\psi s}^{\Lambda}(4459)$ are both $8_1$ flavor pentaquark states and take $m_{P_{\psi}^N}=4440$ MeV and $m_{P_{\psi s}^{\Lambda}}=4459$ MeV as inputs. The three LECs can't be fully determined with only two inputs, and additional constraints are needed.

As $m_s>m_{u,d}$, we easily obtain $m_{P_{\psi s}^{\Sigma}}$<$m_{P_{\psi ss}^{N}}$. Considering the chromomagnetic spin-spin interaction in the quark model, the baryon total mass can be approximated as \(M = \sum m_{\text{quark}} + \Delta M\), where $\Delta M$ is the additional mass from the chromomagnetic interaction. Since \(\Sigma\) baryons contain two identical-flavor quarks with parallel spins, the chromomagnetic contribution \(\Delta M\) is larger than \(\Lambda\) baryons. The final manifestation is $m_{P_{\psi s}^{\Sigma}}$>$m_{P_{\psi s}^{\Lambda}}$. 

\begin{figure}[htbp!]
	\centering
	\includegraphics[width=1.0\linewidth]{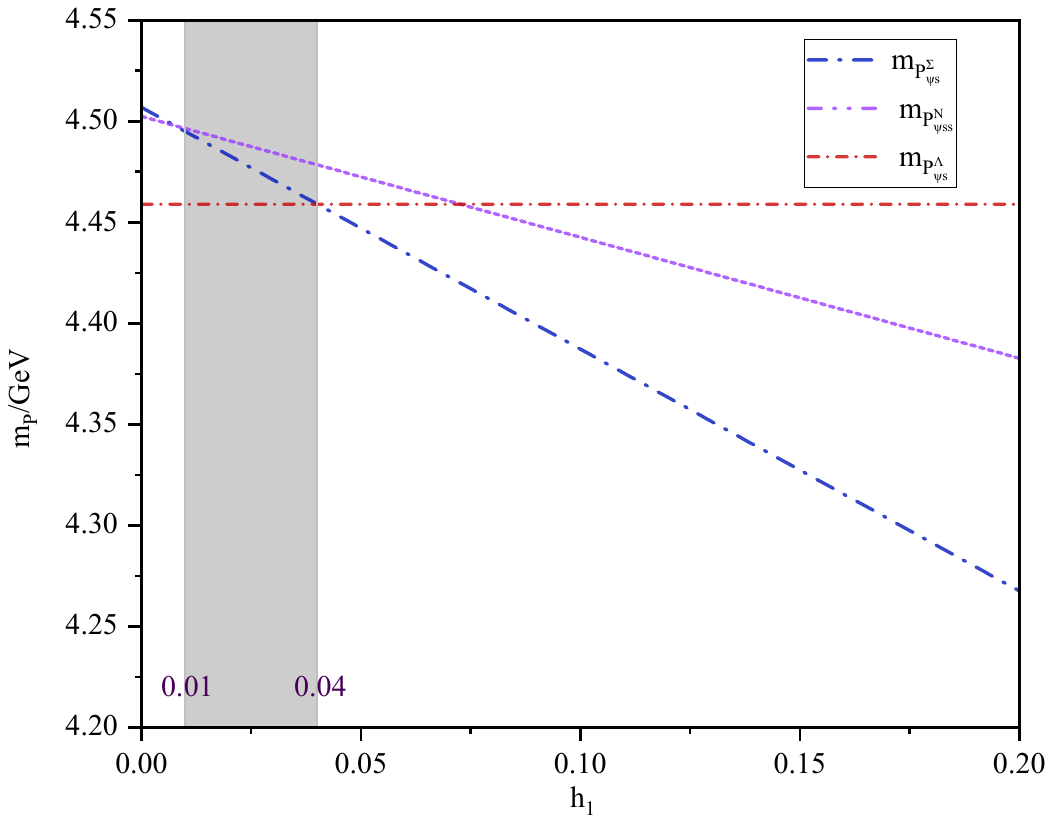}
	\caption{Image of the octet hidden-charm pentaquark state mass as a function of $h_1$. The gray area indicates the value range for $h_1$.}
	\label{fig_2}
\end{figure}

\begin{figure}[htbp]
	\centering
	\includegraphics[width=1.0\linewidth]{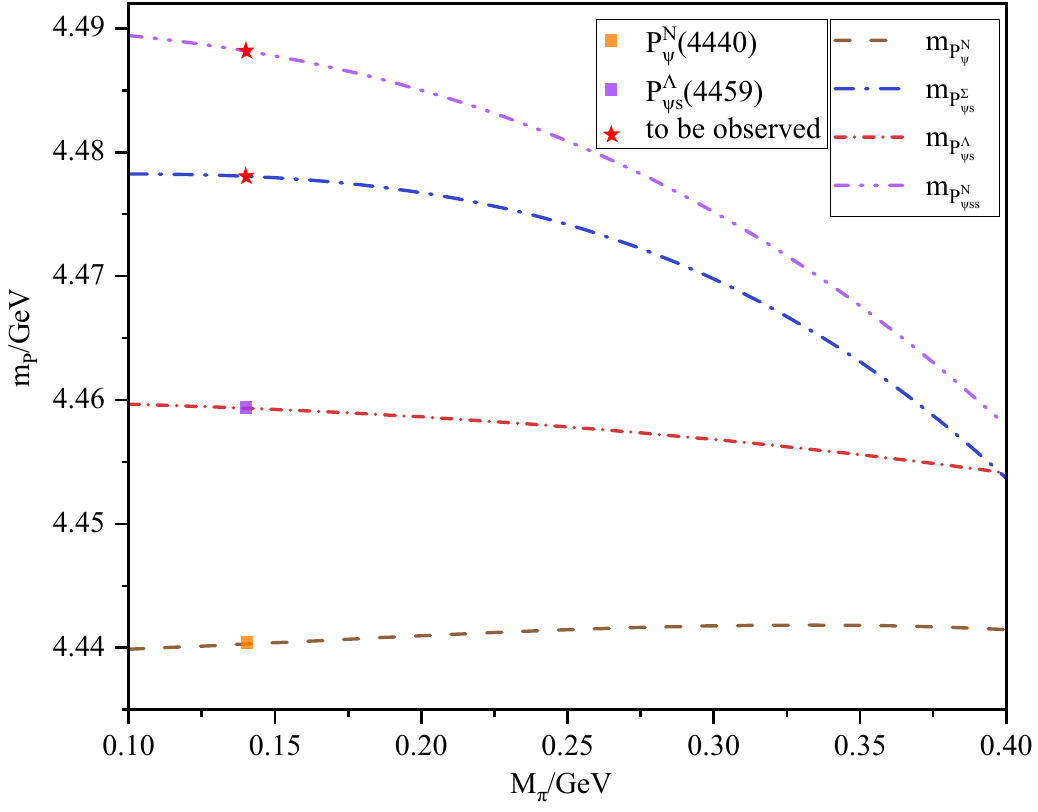}
	\caption{Variation of the octet hidden-charm pentaquark state mass with $M_{\pi}$. The red star and the black box are both physical points when $M_{\pi}=0.140$ GeV, while the red stars are the spin-$\frac{3}{2}$ octet hidden-charm pentaquark states we predict.}
	\label{fig_3}
\end{figure}

With experimental inputs and constraints, we show the pentaquark state mass as a function of $h_1$ in Fig.~\ref{fig_2}, where $h_1$ takes the range 	
\begin{eqnarray}
	0.01< h_1<0.04.
\end{eqnarray}
We take $h_1=0.02$, $h_2=0.02$ and $m_0=4.510$	GeV. Hence the masses of the spin-$\frac{3}{2}$ $8_1$ flavor pentaquarks $m_{P_{\psi s}^{\Sigma}}$ and $m_{P_{\psi ss}^{N}}$ are
	\begin{eqnarray}
	m_{P_{\psi s}^{\Sigma}}&=&4.483\mathrm{GeV},\\
	m_{P_{\psi ss}^{N}}&=&4.490\mathrm{GeV}.
\end{eqnarray}
In Fig.~\ref{fig_3}, we show the dependence of the $8_1$ flavor $P_{\psi}^N$, $P_{\psi s}^{\Sigma}$, $P_{\psi s}^{\Lambda}$, and $P_{\psi ss}^{N}$ pentaquark masses on $M_{\pi}$.
We note that except for $P_{\psi}^N$, all state masses decrease monotonically with increasing \(M_\pi\). Fig.~\ref{fig_3} provides a basis for the chiral extrapolation in lattice QCD and a theoretical guide for the experimental observation of the new pentaquark states.

The $8_2$ flavor hidden-charm pentaquark state mass expression is similar to the $8_1$ flavor one, with only replacement of the corresponding LECs. When we assume that $P_{\psi}^N(4440)$ and $P_{\psi s}^{\Lambda}(4459)$ are both spin-$\frac{3}{2}$ $8_2$ flavor hidden-charm pentaquark states, which similarly yields
\begin{eqnarray}
	-0.02< h_4<0.01.
\end{eqnarray}
We take $h_4=-0.01$, $h_5=0.05$ and $m_0=4.518$ GeV. Thus, the masses of $8_2$ flavor spin-$\frac{3}{2}$ $P_{\psi s}^{\Sigma}$ and $P_{\psi ss}^{N}$ are 
	\begin{eqnarray}
	m_{P_{\psi s}^{\Sigma}}&=&4.475\mathrm{GeV},\\
	m_{P_{\psi ss}^{N}}&=&4.486\mathrm{GeV}.
\end{eqnarray}

In short summary, within the HPChPT framework, we investigate the octet hidden-charm pentaquark state masses with $J^P=\frac{3}{2}^-$ up to NLO. Taking $P_{\psi}^N(4440)$ and $P_{\psi s}^{\Lambda}(4459)$ (with $J^P=\frac{3}{2}^-$) as inputs, we predict that in the $8_1$ flavor, $m_{P_{\psi s}^{\Sigma}}=4.483$ GeV and $m_{P_{\psi ss}^{N}}=4.490$ GeV; in the $8_2$ flavor, $m_{P_{\psi s}^{\Sigma}}=4.475$ GeV and $m_{P_{\psi ss}^{N}}=4.486$ GeV. Predictions for $P_{\psi s}^{\Sigma}$ and $P_{\psi ss}^{N}$ masses help future studies of observables related to octet hidden-charm pentaquark states, as mass is fundamental to nearly all observable expressions. These mass predictions not only refine the energy spectrum picture of the hidden-charm pentaquark but also have profound implications for understanding the dynamics of the strong interaction---specifically, interactions between chiral symmetry breaking, flavor mixing, and multi-quark binding can be revealed. At the same time, these analytic expressions contribute to the chiral extrapolation of lattice QCD. Our results can help further search for $P_{\psi s}^{\Sigma}$ and $P_{\psi ss}^{N}$ in experiments. We anticipate more progress on hidden-charm pentaquark states both theoretically and experimentally.

 \section*{Acknowledgement}
This project is supported by the National Natural Science Foundation of China under Grants No. 11905171. This work is also supported by Natural Science Basic Research Program of Shaanxi (Program No. 2025JC-YBMS-039) and Young Talent Fund of Xi'an Association for Science and Technology (Grant No. 959202413087).


\begin{thebibliography}{99}
		\bibitem{Brambilla:2010cs}
		N.~Brambilla, S.~Eidelman, B.~K.~Heltsley, R.~Vogt, G.~T.~Bodwin, E.~Eichten, A.~D.~Frawley, A.~B.~Meyer, R.~E.~Mitchell and V.~Papadimitriou, \textit{et al.}
		Eur. Phys. J. C \textbf{71} (2011), 1534
	
		\bibitem{Esposito:2016noz}
		A.~Esposito, A.~Pilloni and A.~D.~Polosa,
		Phys. Rept. \textbf{668} (2017), 1-97
		\bibitem{Ali:2017jda}
		A.~Ali, J.~S.~Lange and S.~Stone,
		Prog. Part. Nucl. Phys. \textbf{97} (2017), 123-198
		\bibitem{LHCb:2015yax}
		R.~Aaij \textit{et al.} [LHCb],
		Phys. Rev. Lett. \textbf{115} (2015), 072001
		\bibitem{LHCb:2019kea}
		R.~Aaij \textit{et al.} [LHCb],
		Phys. Rev. Lett. \textbf{122} (2019) no.22, 222001
		\bibitem{Chen:2019asm}
		R.~Chen, Z.~F.~Sun, X.~Liu and S.~L.~Zhu,
		Phys. Rev. D \textbf{100} (2019) no.1, 011502
		\bibitem{Xiao:2019aya}
		C.~W.~Xiao, J.~Nieves and E.~Oset,
		Phys. Rev. D \textbf{100} (2019) no.1, 014021
		\bibitem{Xiao:2019mvs}
		C.~J.~Xiao, Y.~Huang, Y.~B.~Dong, L.~S.~Geng and D.~Y.~Chen,
		Phys. Rev. D \textbf{100} (2019) no.1, 014022
		\bibitem{Guo:2019fdo}
		F.~K.~Guo, H.~J.~Jing, U.~G.~Mei\ss{}ner and S.~Sakai,
		Phys. Rev. D \textbf{99} (2019) no.9, 091501
		\bibitem{Guo:2019kdc}
		Z.~H.~Guo and J.~A.~Oller,
		Phys. Lett. B \textbf{793} (2019), 144-149
		\bibitem{Burns:2019iih}
		T.~J.~Burns and E.~S.~Swanson,
		Phys. Rev. D \textbf{100} (2019) no.11, 114033
		\bibitem{Wang:2019spc}
		G.~J.~Wang, L.~Y.~Xiao, R.~Chen, X.~H.~Liu, X.~Liu and S.~L.~Zhu,
		Phys. Rev. D \textbf{102} (2020) no.3, 036012
		\bibitem{Wang:2019ato}
		B.~Wang, L.~Meng and S.~L.~Zhu,
		JHEP \textbf{11} (2019), 108
	\bibitem{He:2019ify}
	J.~He,
	Eur. Phys. J. C \textbf{79} (2019) no.5, 393
	\bibitem{He:2019rva}
	J.~He and D.~Y.~Chen,
	Eur. Phys. J. C \textbf{79} (2019) no.11, 887
	\bibitem{Lin:2019qiv}
	Y.~H.~Lin and B.~S.~Zou,
	Phys. Rev. D \textbf{100} (2019) no.5, 056005
	\bibitem{Yalikun:2021bfm}
	N.~Yalikun, Y.~H.~Lin, F.~K.~Guo, Y.~Kamiya and B.~S.~Zou,
	Phys. Rev. D \textbf{104} (2021) no.9, 094039
	\bibitem{PavonValderrama:2019nbk}
	M.~Pavon Valderrama,
	Phys. Rev. D \textbf{100} (2019) no.9, 094028
	\bibitem{Du:2021fmf}
	M.~L.~Du, V.~Baru, F.~K.~Guo, C.~Hanhart, U.~G.~Mei\ss{}ner, J.~A.~Oller and Q.~Wang,
	JHEP \textbf{08} (2021), 157
	
	\bibitem{Peng:2024yzn}
	F.~Z.~Peng, L.~S.~Geng and J.~J.~Xie,
	Phys. Rev. D \textbf{111} (2025) no.5, 054029
	\bibitem{LHCb:2021chn}
	R.~Aaij \textit{et al.} [LHCb],
	Phys. Rev. Lett. \textbf{128} (2022) no.6, 062001
	\bibitem{Shen:2017ayv}
	C.~W.~Shen, D.~R\"onchen, U.~G.~Mei\ss{}ner and B.~S.~Zou,
	Chin. Phys. C \textbf{42} (2018) no.2, 023106
		\bibitem{LHCb:2020jpq}
		R.~Aaij \textit{et al.} [LHCb],
		Sci. Bull. \textbf{66} (2021), 1278-1287
		\bibitem{Zou:2021sha}
		B.~S.~Zou,
		Sci. Bull. \textbf{66} (2021), 1258
		\bibitem{Karliner:2021xnq}
		M.~Karliner and J.~L.~Rosner,
		Sci. Bull. \textbf{66} (2021) no.13, 1256
		\bibitem{Peng:2020hql}
		F.~Z.~Peng, M.~J.~Yan, M.~S\'anchez S\'anchez and M.~P.~Valderrama,
		Eur. Phys. J. C \textbf{81} (2021) no.7, 666
		\bibitem{Chen:2020uif}
		H.~X.~Chen, W.~Chen, X.~Liu and X.~H.~Liu,
		Eur. Phys. J. C \textbf{81} (2021) no.5, 409
		\bibitem{Chen:2020kco}
		R.~Chen,
		Phys. Rev. D \textbf{103} (2021) no.5, 054007
		\bibitem{Lu:2021irg}
		J.~X.~Lu, M.~Z.~Liu, R.~X.~Shi and L.~S.~Geng,
		Phys. Rev. D \textbf{104} (2021) no.3, 034022
		\bibitem{LHCb:2022ogu}
		R.~Aaij \textit{et al.} [LHCb],
		Phys. Rev. Lett. \textbf{131} (2023) no.3, 031901
		\bibitem{Ali:2019npk}
		A.~Ali and A.~Y.~Parkhomenko,
		Phys. Lett. B \textbf{793} (2019), 365-371
		\bibitem{Wang:2019got}
		Z.~G.~Wang,
		Int. J. Mod. Phys. A \textbf{35} (2020) no.01, 2050003
		\bibitem{Cheng:2019obk}
		J.~B.~Cheng and Y.~R.~Liu,
		Phys. Rev. D \textbf{100} (2019) no.5, 054002
		\bibitem{Zhu:2019iwm}
		R.~Zhu, X.~Liu, H.~Huang and C.~F.~Qiao,
		Phys. Lett. B \textbf{797} (2019), 134869
	\bibitem{Pimikov:2019dyr}
	A.~Pimikov, H.~J.~Lee and P.~Zhang,
	Phys. Rev. D \textbf{101} (2020) no.1, 014002
	\bibitem{Ruangyoo:2021aoi}
	W.~Ruangyoo, K.~Phumphan, C.~C.~Chen, A.~Limphirat and Y.~Yan,
	J. Phys. G \textbf{49} (2022) no.7, 075001
	\bibitem{Fernandez-Ramirez:2019koa}
	C.~Fern\'andez-Ram\'\i{}rez \textit{et al.} [JPAC],
	Phys. Rev. Lett. \textbf{123} (2019) no.9, 092001
	\bibitem{Nakamura:2021qvy}
	S.~X.~Nakamura,
	Phys. Rev. D \textbf{103} (2021), 111503
	\bibitem{Burns:2022uiv}
	T.~J.~Burns and E.~S.~Swanson,
	Phys. Rev. D \textbf{106} (2022) no.5, 054029

	\bibitem{Weinberg:1978kz}
	S.~Weinberg,
	Physica A \textbf{96} (1979) no.1-2, 327-340
	\bibitem{Gasser:1983yg}
	J.~Gasser and H.~Leutwyler,
	Annals Phys. \textbf{158} (1984), 142
	\bibitem{Scherer:2012zzd}
	S.~Scherer and M.~R.~Schindler,
	Lect. Notes Phys. \textbf{830} (2012), 1-48
	\bibitem{Huang:2019not}
	B.~L.~Huang and J.~Ou-Yang,
	Phys. Rev. D \textbf{101} (2020) no.5, 056021
	\bibitem{Li:2024jlq}
	H.~S.~Li,
	Phys. Rev. D \textbf{109} (2024) no.11, 114039
	\bibitem{Li:2025hpd}
	H.~S.~Li and T.~Li,
	[arXiv:2502.05495 [hep-ph]].
	\bibitem{Sun:2014aya}
	Z.~F.~Sun, Z.~W.~Liu, X.~Liu and S.~L.~Zhu,
	Phys. Rev. D \textbf{91} (2015) no.9, 094030
	\bibitem{Jenkins:1990jv}
	E.~E.~Jenkins and A.~V.~Manohar,
	Phys. Lett. B \textbf{255} (1991), 558-562
	\bibitem{Jenkins:1992pi}
	E.~E.~Jenkins, M.~E.~Luke, A.~V.~Manohar and M.~J.~Savage,
	Phys. Lett. B \textbf{302} (1993), 482-490
	[erratum: Phys. Lett. B \textbf{388} (1996), 866-866]
	\bibitem{Yamaguchi:2019seo}
	Y.~Yamaguchi, H.~Garc\'\i{}a-Tecocoatzi, A.~Giachino, A.~Hosaka, E.~Santopinto, S.~Takeuchi and M.~Takizawa,
	Phys. Rev. D \textbf{101} (2020) no.9, 091502
		\bibitem{Liu:2019zvb}
		M.~Z.~Liu, T.~W.~Wu, M.~S\'anchez S\'anchez, M.~P.~Valderrama, L.~S.~Geng and J.~J.~Xie,
		Phys. Rev. D \textbf{103} (2021) no.5, 054004
\bibitem{Chen:2021tip}
R.~Chen,
Eur. Phys. J. C \textbf{81} (2021) no.2, 122
	\bibitem{Wang:2022gfb}
X.~W.~Wang and Z.~G.~Wang,
Int. J. Mod. Phys. A \textbf{37} (2022) no.31n32, 2250189
\bibitem{Ozdem:2024rch}
U.~\"Ozdem,
Phys. Rev. D \textbf{111} (2025) no.7, 074038
\bibitem{Mutuk:2024ltc}
H.~Mutuk and X.~W.~Kang,
Phys. Lett. B \textbf{855} (2024), 138772
\bibitem{Ozdem:2022kei}
U.~\"Ozdem,
Phys. Lett. B \textbf{836} (2023), 137635
\bibitem{Li:2024wxr}
H.~S.~Li, F.~Guo, Y.~D.~Lei and F.~Gao,
Phys. Rev. D \textbf{109} (2024) no.9, 094027
\bibitem{Severt:2019sbz}
D.~Severt, U.~G.~Mei\ss{}ner and J.~Gegelia,
JHEP \textbf{03} (2019), 202

		\end{thebibliography}
			\end{document}